\documentclass[doublecol]{epl2m} 

\usepackage{amsmath}
\usepackage{latexsym}
\usepackage{amsfonts}
\usepackage{amssymb}

\usepackage{bbm,dsfont}
\usepackage{dsfont}
\usepackage{graphicx}
\usepackage{hyperref}

\usepackage{verbatim}

\newtheorem{proposition?}{Proposition?}

\usepackage{color}

\usepackage[normalem]{ulem} 

\newcommand{\mods}[1]{\left \vert #1 \right \vert ^2}

\newcommand{\hi}{\mathcal{H}} 
\newcommand{\his}{\mathcal{H}_{\mathcal{S}}}
\newcommand{\hir}{\mathcal{H}_{\mathcal{R}}}

\newcommand{\hit}{\mathcal{H}_{\mathcal{T}}}
\newcommand{\Y}{\yen}

\newcommand{\lh}{\mathcal{L(H)}} 
\newcommand{\lhs}{\mathcal{L}(\mathcal{H}_{\mathcal{S}})} 
\newcommand{\lht}{\mathcal{L}(\mathcal{H}_{\mathcal{T}})} 

\newcommand{\lhr}{\mathcal{L}(\hir)} 

\newcommand{\ip}[2]{\left\langle\,#1\,{|}\,#2\,\right\rangle} 
\newcommand{\ket}[1]{|#1\rangle} 
\newcommand{\bra}[1]{\langle#1|} 
 
\newcommand{\tr}[1]{\textrm{tr}\left[#1\right]} 
\newcommand{\id}{\mathbbm{1}} 

\newcommand{\Esf}{\mathsf{E}}

\newcommand{\Fsf}{\mathsf{F}}

\newcommand{\Sy}{\mathcal{S}}

\newcommand{\R}{\mathcal{R}}
\newcommand{\T}{\mathcal{T}}

\newcommand{\E}{\mathsf{E}}

\title{Relativity of Quantum States and Observables}
\shorttitle{Relativity of Quantum States and Observables} 

\subtitle{\normalsize Published in: EPL {\bf 117} (2017) 40004, DOI: \href{http://dx.doi.org/10.1209/0295-5075/117/40004}{10.1209/0295-5075/117/40004}}

\author{
L. Loveridge\inst{1} \and P. Busch\inst{2} \and T. Miyadera\inst{3}}
\shortauthor{L. Loveridge \etal}

\institute{                    
  \inst{1} Descartes Centre for the History and Philosophy of Science and the Humanities and Department of Mathematics, Utrecht University, 3584 CC Utrecht,  The Netherlands\\
  \inst{2} Department of Mathematics, University of York, Heslington, York YO10 5DD, UK\\
  \inst{3} Department of Nuclear Engineering, Kyoto University, 
Kyoto daigaku-katsura, Nishikyo-ku, Kyoto, Japan 615-8540
}


\abstract{Under the principle that quantum mechanical observables are invariant under
relevant symmetry transformations, we explore how the usual, non-invariant quantities
may capture measurement statistics. Using a relativisation mapping, viewed
as the incorporation of a quantum reference frame, we show that the usual quantum description approximates the relative one precisely when the reference system admits an appropriate localisable quantity and a localised state. From this follows a new perspective on the nature and reality of
quantum superpositions and optical coherence.}

\begin{document}

\maketitle

\section{Introduction}
The Hilbert space formulation of quantum theory provides an empirically successful 
theoretical account of laboratory experiments. However, as recognised soon after the inception  of this  theory \cite{bor1, ed1}, and revisited at various times throughout its development ({\it e.g.}, \cite{as, ak1, ak2, sp1, brs}), attention must be paid to the fact that quantum observables are defined and measured relative to a reference frame---they are {\em relational} 
attributes.\footnote{This language is reminiscent of work presented in, {\it e.g.}, 
C. Rovelli, Relational quantum mechanics, Int.~J.~Theor.~Phys.~{\bf 35}, 1637 (1996), and G. Bene and D. Dieks, A perspectival version of the modal interpretation of quantum mechanics and the origin of macroscopic behavior, Found.~Phys.~ {\bf 32}, 5 (2002). Though we see no {\it prima facie} formal connection between our notion of relationalism and theirs, we think it worth investigating further.} The fact that such reference frames are themselves quantum systems poses challenges concerning their precise definition and interpretation.
Thus it is paramount to understand the features of quantum reference systems that allow them to 
properly fulfil their role.

There is an abundance of literature on quantum reference frames. This includes work of mainly foundational interest, for example, the possibility of extending the relativity principle to quantum mechanics \cite{ak2,pb1} and the ``view" from a quantum frame of reference \cite{sp1}, the possibility of practically obviating superselection rules \cite{www,as,www70,strowi,brs}, and the role played by reference frames in the reality of optical coherence \cite{molmer, dia, brs}. On the more practical side, work on reference frames has been phrased to a large extent within the framework of resource theories (including asymmetry and coherence, {\it e.g.}, \cite{brs,ms1,ajr1,pian1,pian2}), used in a range of applications in quantum information science.

Previous work on the subject of quantum reference frames has lacked mathematical precision
and has not provided a coherent conceptual account of the relationship between the description with (``relative/relational") and without (``absolute") reference to a frame. In this paper, we provide a rigorous mathematical apparatus from which, under the physical principle that all measurable quantities are invariant under given symmetry transformations ({\it e.g.}, shifts, rotations, {\it etc.}), a resolution of a range of conceptual problems follows naturally. 

The relativisation procedure we provide (eq. \eqref{eq:yen}) generalises (by considering positive operator valued measures for reference quantities) and makes rigorous a map in \cite{brs}; we use it to construct invariant quantities out of arbitrary (``absolute") ones, and compare the two descriptions. Arguing that
``absolute" quantities 
are mere theoretical symbols, not represented in reality, but corresponding to relative quantities
for which the frame-dependence has been suppressed, we show that accurate representation of relative quantities follows from the presence of an appropriately \emph{localised} reference quantity. This, we argue, has a clear physical interpretation but has thus far been absent from the reference frames literature. 
Focussing on the case of $U(1)$ symmetries associated with phase-like quantities, we use our scheme to investigate the meaning of quantum coherence---optical coherence in particular---which we show to 
be a relational notion, and introduce the concept of mutual coherence between two systems. This allows for a completely relational approach to quantum coherence. We point out where our perspective differs from the mainstream view throughout the text; in \cite{lbm1} we provide detailed analysis and examples; here we present 
 a summary of our main
findings.

\section{Preliminaries}
Associated to each physical system
is a complex Hilbert space  $\hi$. The space of bounded operators on $\hi $ is denoted
$\lh$ and the trace class $\mathcal{L}_1(\hi)$, the positive trace-one elements of which
we identify with the states, the pure states being singled out as the rank one projections of the form $P[\phi] \equiv \ket{\phi}\bra{\phi}$ for some unit vector $\phi$. An observable is represented by a positive operator measure ({\sc pom}) $\Esf: \mathcal{F} \to \lh$ on  
a $\sigma$-algebra $\mathcal{F}$ of subsets of the outcome set $\Omega$, i.e., ${0}\leq \Esf(X) \leq \id$, $\Esf(\Omega)=\id$, and $\Esf$ is 
($\sigma$-)additive on disjoint sets. 
If $\Esf$ is a projection valued measure,  
the observable represented by $\E$ is called {\em sharp} and is the spectral measure of a unique self-adjoint operator $A$. $\E$ will be called {\em unsharp} otherwise. $\mathcal{F}$ is typically identified in this paper with the Borel
sets $\mathcal{B}(\Omega)$  of the real line ($\Omega=\mathbb{R}$) or the circle ($\Omega=S^1$), 

The system part of a quantum system-plus-reference will be denoted $\Sy$ with Hilbert space $\his$, 
the reference system $\R$ has Hilbert space $\hir$, and the compound system $\Sy + \R$ has Hilbert space $\hi_{\T} =\his \otimes \hir$.

\section{Symmetry, Relativisation, and Restriction}

{\em Phase shift invariance.}
Consider the circle group $S^1$, which we identify with the interval $[-\pi, \pi]$ (identifying also $-\pi$ and $\pi$)
and continuous unitary representations $U_{\Sy}$ and $U_{\R}$, with $U_{\Sy}(\theta) = e^{iN_{\Sy}\theta}$ and
$U_{\R}(\theta) = e^{iN_{\R}\theta}$, acting in $\lhs$ and $\lhr$ respectively.
The tensor product representation is written $U=U_{\Sy}\otimes U_{\R}$ with generator $N_{\T}:=N_{\Sy}+N_{\R}$. Here, $N_{\Sy}$ and $N_{\R}$ are number operators of the form $N=\sum_{\mathcal{N}}nP_n$ with $P_n$ possibly degenerate
projections and $\mathcal{N}$ a possibly infinite collection of integers.

``Absolute" phase is characterised by phase-shift covariance (see eq. \eqref{eq:pscov}), and such a quantity
naturally arises in (for example) the study of laser light. However, what is actually measured is a relative phase between two lasers, which is a phase-shift invariant quantity. For a system $\Sy$ in isolation, measurable quantities of $\Sy$ must satisfy $U_{\Sy}(\theta)\Esf(X)U_{\Sy}(\theta)^*=\Esf(X)$
for all $X \in \mathcal{B}(S^1)$ (equivalently, $[\Esf(X),N_{\Sy}] = 0$ for all $X$). If, instead,
we consider $\Sy + \R$, observables must be invariant under $U(\theta)$, or equivalently commute with $N_{\T}$. This then opens the possibility
of observables (as invariant quantities) of $\Sy + \R$ being (possibly approximately) described by non-invariant 
quantities of $\Sy$; the meaning and adequacy of such a description is of central concern in this paper.

\subsection{Symmetrisation} A phase-shift invariant sharp observable $A\in \lhs$ satisfies
\begin{equation*}
A = \sum_{\mathcal{N}}P_n A P_n =: \tau_{\Sy}(A).
\end{equation*}
We write $\tau_{\R}$ and $\tau$ for the corresponding maps on $\lhr$ and 
$\lht$, respectively. The following holds: 
\begin{equation}\label{eq:taus}
\tr{\tau_{\Sy}(A)\rho} = \tr{A {\tau_{\Sy}}{_*}(\rho)} = \tr{\tau_{\Sy}(A){\tau_{\Sy}}{_*}(\rho)}
\end{equation}
for all $A \in \lhs$, $\rho \in \mathcal{L}_1(\his)$. The first equality defines 
${\tau_{\Sy}}{_*}: \mathcal{L}_1(\his) \to \mathcal{L}_1 (\his)$ as the predual to $\tau_{\Sy}$;
it is unique and trace-preserving, and has the same form as $\tau_{\Sy}$. The invariance of a density matrix $\rho \in \mathcal{L}_1(\his)$ under $U_{\Sy}$ is equivalent to
its ${\tau_{\Sy}}{_*}$-invariance; Eq.~\eqref{eq:taus} entails that, provided one only considers invariant quantities, $\rho$ and ${\tau_{\Sy}}_* (\rho)$ are observationally equivalent.
Dually, if one uses only invariant states, $A$ and $\tau_{\Sy}(A)$ cannot be distinguished.
These observations also hold, {\it mutatis mutandis}, for $\R$ and $\Sy + \R$.

This has immediate impact on the possibility
of observing coherence of superpositions: for example, if only the system $\Sy$ is taken in isolation, 
the superposition state $\Psi = \sum_n c_n \ket{n}$
 cannot be distinguished by any invariant quantity from the mixed state
${\tau_{\Sy}}_*(P[\Psi]) = \sum_n\mods{c_n}\ket{n}\bra{n}$. Since $\Psi$ could be a coherent 
state ($c_n = e^{-\mods{\beta}/2}\beta^n / \sqrt{n!}$), whether there is any empirical difference between representing the state of a laser by $P[\Psi]$ rather than ${\tau_{\Sy}}_*(P[\Psi])$ has profound implications upon whether the output of a laser is ``really" coherent.

\subsection{Relativisation}
Let $\Fsf: \mathcal{B}(S^1) \to \lhr$ be a {\sc pom} with the covariance property
\begin{equation}\label{eq:pscov}
e^{iN_{\R}\theta}\Fsf(X) e^{-iN_{\R}\theta} = \Fsf(X + \theta)
\end{equation}
(addition modulo $2 \pi$). $N_{\R}$ is a number observable for $\R$ and $\Fsf$ is called
a (covariant) phase {\sc pom}; we note that if the spectrum of $N_{\R}$ is bounded from below,
as in the harmonic oscillator Hamiltonian for example, $\Fsf$ is never projection valued \cite{hol1, ljp1}. 

We define a ``relativising" map
$\Y: \lhs \to \lht$ by 
\begin{equation}\label{eq:yen}
\Y(A) = \int_{S^1}U_{\Sy}(\theta)AU_{\Sy}(\theta)^*\otimes \Fsf(d \theta)
\end{equation}
which takes ``absolute" quantities to relative ones---$\Y(A)$ is invariant under the representation $U=U_{\Sy}\otimes U_{\R}$ of $S^1$ (i.e., $\tau(\Y(A))=\Y(A)$). The construction of $\Y$ requires the theory of integration of operator-valued functions with respect to an operator measure; following \cite{mlak} we may
first define $\Y$ on a suitable dense subset of $\mathcal{L}(\his)$, extended to $\lhs$ as described in \cite{lbm1}.

We find that $\Y$ is bounded, normal (ensuring the existence of a unique predual map which acts on states) and completely positive \cite{lbm1}, and $\Y(\tau_{\Sy}(A)) = \tau_{\Sy}(A) \otimes \id$. $\Y$ is a rigorous and more general version of the ``$\$$" map appearing in \cite{brs}. We note that $\Y$ does not act on states, as was mistakenly assumed for the $\$$ there, and instead
one must consider its predual $\Y_{*}$ and the inverse image $\Y_* ^{-1} (\{ \cdot\})$. $\Y$ also differs from $\$$ in that there is no recourse to improper eigenstates of continuous-spectrum operators, and the relativising quantity $\Fsf$ may be unsharp.

We will hereafter assume that $\Fsf$ has the localisation property known as the {\it norm-$1$} property \cite{tnorm1}---for any
$X$ with $\Fsf(X)\neq 0$ there is a sequence of unit vectors 
$(\phi_i)$ in $\hir$ for which $\lim_{i \to \infty}\ip{\phi_i}{\Fsf (X) \phi_i} = 1$. For this property to hold, the Hilbert space $\hi_{R}$ must be 
infinite dimensional \cite{mlbshort}.
All projection valued observables
have such a property, and in this special case $\Y$ is a $^*$-homomorphism and therefore preserves the algebraic structure of $\lhs$.

$\Y$ has the effect of relativising given self-adjoint operators, and more generally {\sc pom}s, of $\his$. If $\Fsf_{\Sy}$ is a covariant phase {\sc pom}, $\Y \circ \Fsf_{\Sy}$ is a relative phase observable \cite{jpthesis, lbm1}. If $\Fsf = \Esf^{\Phi _{\R}}$ is the spectral measure
of the self-adjoint azimuthal angle operator ${\Phi}_{\R}$ conjugate to angular momentum along some axis, and  $\Phi_{\Sy}$ is the analogous quantity of $\Sy$, then $\Y(\Phi_{\Sy}) = \Phi_{\Sy} - \Phi_{\R}$: the relative angle between the system and reference. $\Y$ may also be defined with respect to the shift group on $\mathbb{R}$ by replacing in \eqref{eq:yen} $U_{\Sy}(\theta)$ with $e^{ixP}$ ($P$ the momentum) and $\Fsf$ with the spectral measure of the position $Q_{\R}$. Then  $\Y(Q_{\Sy}) = Q_{\Sy} - Q_{\R}$: the relative position. We view the ``absolute" operators as formally representing their relative counterparts;
as well as relativising the familiar quantities just discussed, $\Y$ also relativises any ``absolute" quantity represented by an arbitrary {\sc pom}. The important question of the adequacy of such an absolute description as approximating the relative/invariant one shall be addressed shortly.

The predual map
 $\Y_*:\mathcal{L}_1(\hit) \to \mathcal{L}_1(\his)$ may be written explicitly on  product states as 
\begin{equation*}
\Y_* (\rho_{\Sy} \otimes \rho_{\R}) = \int_{S^1}U_{\Sy}(\theta)^*\rho_{\Sy}U_{\Sy}(\theta) \tr{\rho_{\R}\Fsf(d \theta)}
\end{equation*}
and extended to all of $\mathcal{L}_1(\hi) $ by linearity and continuity. Just as $\Y$ relativises observables, $\Y_*$ ``derelativises" states. In fact,
it is a precise version of a ``dequantization" map of \cite{brs}.

\subsection{Restriction}
Consider now the isometric embedding $\mathcal{V}_{\omega} :\mathcal{L}_1(\his) \to \mathcal{L}_1(\hit)$ defined by $\rho \mapsto \rho \otimes \omega$. This has a dual (\emph{restriction}) map $\Gamma_{\omega}:\lht \to \lhs$, which on tensor product operators $A \otimes B$ takes the form 
\begin{equation*}
\Gamma_{\omega}(A \otimes B) = A~ \tr{\omega B},
\end{equation*}
extended by linearity and continuity to $\lht$. $\Gamma_{\omega}$ is a channel, restricts {\sc pom}s of $\Sy + \R$ to those of $\Sy$;
and is used to translate back from the relative picture to the ``absolute" one, contingent upon the state $\omega$ of $\R$. 

We may compose $\Gamma_{\omega}$ with $\Y$ which yields the map $\lhs \to \lhs$:
\begin{equation}\label{eq:yenres}
(\Gamma_{\omega}\circ \Y)(A) = \int_{S^1}U_{\Sy}(\theta)AU_{\Sy}(\theta)^* \mu_{\omega}^{\Fsf}(d \theta),
\end{equation}
where the measure $\mu_{\omega}^{\Fsf}(X):=\tr{\omega\Fsf(X)}$ and we observe that $(\Gamma_{\omega} \circ \Y)_* = \Y_* \circ \mathcal{V}_{\omega}$. We investigate the agreement between $A$ and $(\Gamma_{\omega} \circ \Y)(A)$ and discuss the consequences.

\section{Reference Localisation and Coherence}

In the previous section we described how $\Y$ relativises arbitrary ``absolute" quantities of $\Sy$,
giving invariant quantities of $\Sy + \R$,
and how the restriction map $\Gamma_{\omega}$ recovers a description in terms of $\Sy$ alone.
The localisation properties of the measure $\mu_{\omega}^{\Fsf}$ 
dictate the quality of the approximation of $(\Gamma_{\omega}\circ \Y)(A)$ by $A$. We now consider 
the extremes: high reference state localisation and complete delocalisation, and discuss the implications.

\subsection{Localisation}
Consider a sequence of unit vectors $(\phi_i)\subset \hir$ which becomes increasingly well
localised around $0$ with respect to a covariant phase {\sc pom} $\Fsf$ (which satisfies 
the norm-$1$ property). Then,  
\begin{equation}\label{eq:lim}
\lim_{i \to \infty}(\Gamma_{\phi_i}\circ \Y)(A) = A 
\end{equation}
in the topology of pointwise convergence of 
expectation values \cite{lbm1}.\footnote{As shown in \cite{mlbshort}, the convergence holds in the operator norm topology in case $[N_{\Sy}, A]$ is bounded.} Intuitively, as $\mu_{\phi_i}^{\Fsf}$ becomes concentrated around $\theta = 0$, the only contribution to the integral \eqref{eq:yenres} is on a vanishingly small neighbourhood of $0$. For instance, high amplitude coherent states peaked around zero phase are ``near" eigenstates
of $\Fsf^{\rm can}$---the canonical phase \cite{pekju}---with approximate ``eigenvalue" $0$, and the above limit may be equivalently understood as the high-amplitude limit in a set of coherent states.

For any 
$\rho \in \mathcal{L}_1(\his)$ and $A \in \lhs$, 
\begin{equation}\label{eq:yenp}
\lim_{i \to \infty}\tr{\tau_*(\rho \otimes P[\phi_i])\Y(A)} = \tr{\rho A},
\end{equation}
and hence good approximation of $A$ by $\Y(A)$ can be done using only invariant states
of $\Sy + \R$.
 
This proves that if the reference system $\R$ has a localisable phase-like quantity, the description of the system alone is a good approximation of the relative/invariant one, with the quality of approximation being arbitrarily good given sufficiently well localised states of $\R$. Though high amplitude coherent states have been shown 
to be important in these kinds of considerations ({\it e.g.}, \cite{brs}, \cite{dia}, \cite{dbrs}), to our knowledge it has never been pointed out that the high phase localisation is the crucial property.

\subsection{Delocalisation} 
At the other extreme to high localisation, eigenstates of $N_{\R}$ and their mixtures
are completely phase delocalised, in which case the measure in \eqref{eq:yenres} is the Haar measure on $S^1$. Indeed, consider the state ${\tau_{\R}}_*(\rho_{\R})$. Then,
\begin{equation}\label{eq:pdel}
(\Gamma_{{{\tau_{\R}}_*}(\rho_{\R})}\circ \Y)(A) = \frac{1}{2 \pi}\int_{S^1}U_{\Sy}(\theta) A U_{\Sy}(\theta)^* d \theta
\end{equation}
for any $\rho_{\R}$, even highly localised or coherent. On states of $\Sy$, 
$(\Gamma_{{{\tau_{\R}}_*}(\rho_{\R})}\circ \Y)_*: \mathcal{L}_1(\his) \to \mathcal{L}_1(\his)$ takes the form
\begin{equation*}
(\Gamma_{{{\tau_{\R}}_*}(\rho_{\R})}\circ \Y)_*(\rho) = \frac{1}{2 \pi}\int_{S^1} U_{\Sy}(\theta)^*\rho U_{\Sy}(\theta) d \theta.
\end{equation*}
This is the ``twirling" map
appearing in {\it e.g}. \cite{brs}. There, it results from an epistemic restriction when two experimenters do not share a phase reference. Here, if, for instance, $\rho_{\R}$ is a number eigenstate, by the preparation uncertainty relation for number and phase, $\rho_{\R}$ is completely phase indefinite---a quantum restriction. In this case, the statistics for $\Sy$ are never well represented by an arbitrary state and ``absolute" quantity not commuting with $N_{\Sy}$.

\subsection{Coherence} The previous analysis justifies the use of ``absolute" quantities of $\Sy$ along with 
an unrestricted state description whenever the state of $\R$ is highly localised.
An apparent circularity thus arises: in order to speak of ``absolute" quantities and coherent/localised states of $\Sy$ (i.e., those which are not invariant under $\tau_{\Sy *}$) as representing their invariant counterparts of $\Sy + \R$, ``absolute" quantities and coherent/localised states are presumed for $\R$. This situation was highlighted 
by Wick, Wightman and Wigner \cite{www70} in response to \cite{as}, in the context of superselection rules, and reappeared in \cite{dia} in a hypothetical dialogue concerning the reality of coherence in an optical setting. 

However, no such inconsistency arises if one speaks only of coherence and localisation of \emph{pairs} of states of $\Sy$ and $\R$. On the question of coherence, we observe that the following statements are equivalent: (i) there exists
an invariant observable $\Esf$ of $\Sy + \R$ and $X$
such that $\tr{({\tau_{\Sy}}_*(\rho_{\Sy})\otimes \rho_{\R})\Esf(X)} \neq \tr{(\rho_{\Sy}\otimes \rho_{\R})\Esf(X)}$, (ii) 
there exists
an invariant observable $\Esf$ of $\Sy + \R$ and $X$ such that $\tr{(\rho_{\Sy}\otimes {\tau_{\R}}_*(\rho_{\R})\Esf(X)} \neq \tr{(\rho_{\Sy}\otimes \rho_{\R})\Esf(X)}$. Statement (i) means that 
$\rho_{\Sy}$ is coherent relative to $\rho_{\R}$ and the equivalent statement (ii) that $\rho_{\R}$ is coherent relative to $\rho_{\Sy}$. We refer to any pair $(\rho_{\Sy}, \rho_{\R})$ satisfying (i) or (ii) as {\em mutually coherent}. An example is found in the contrasting expressions \eqref{eq:lim} and \eqref{eq:pdel}: by choosing
an invariant quantity of the form $\Y(A)$ in place of the {\sc pom} $\E$ above and $\rho_{\R}$ highly (``absolutely") localised, we may make
$\tr{(\rho_{\Sy} \otimes \rho_{\R})\Y(A)} \approx \tr{\rho_{\Sy}A}$, with as good approximation as one chooses. However, generically, the discrepancy between $\tr{\rho_{\Sy}A}$ and $\tr{(\rho_{\Sy}\otimes {\tau_{\R}}_*(\rho_{\R}))\Y(A)}$ is large.

Just as coherence is a relational notion, depending on both $\Sy$ and $\R$, so is localisation.
Specifically, states with ``absolute" coherence/localisation can be recovered in the high reference phase localisation limit: Eq.~\eqref{eq:yenp} gives 
\begin{equation}\label{eq:ytau}
\lim_{i \to \infty}\Y_{*}(\tau_*(\rho_{\Sy} \otimes P[\phi_i]) = \rho{_{\Sy}}
\end{equation}
in the weak sense, showing that any state of $\Sy$ (possibly highly localised) can be approximated by invariant states of 
$\Sy + \R$. ``Absolute" coherence/localisation of states of $\Sy + \R$ is not required. The usual
reading of localisation of $\rho_{\Sy}$ may be interpreted as referring to the \emph{relational localisation} of $(\rho_{\Sy}, \rho_{\R})$ in $\tau_{*}(\rho_{\Sy} \otimes \rho_{\R})$. This latter state has no ``absolute" coherence or localisation, and therefore does not require recourse to yet another system to find its meaning. Moreover, the state $\tau_{*}(\rho_{\Sy} \otimes \rho_{\R})$
is not to be understood as ``containing" ``absolutely" localised/coherent states, due to the partial trace over either system giving an invariant/delocalised/absolutely incoherent state description.

Just as ``absolute" quantities represent their invariant counterparts, with good approximation in the appropriate
limit, the same may be said for $\rho_{\Sy}$ and the collection $\{\tau_* (\rho_{\Sy} \otimes P[\phi_i])\}$ of relative or ``relational states". 

These observations have clear bearing on the issue of optical coherence of laser beams---a subject of much controversy culminating in \cite{dia}, where the 
relational aspects of quantum coherence was emphasised, but little in the way of a formal framework was provided. In light of the relational character of coherence, the question of whether a laser beam is coherent
on its own is not meaningful. One can instead enquire about the mutual coherence of a system-reference pair, and whether the reduced ``absolute" coherence with an ``absolute" quantity provides an empirically adequate description of the given composite and a relative observable.

Hence we may consider an ``absolute" phase observable $\Fsf^{\Sy}$ and a 
coherent state $\ket{\beta} = \sum_n c_n \ket{n}$ of $\Sy$, and construct a 
relative phase observable 
$\Fsf^T = \Y\circ \Fsf ^{\Sy}$,
so that
\begin{align*}
\ip{\beta}{\Fsf^{\Sy}(X)\beta}
 &= \lim_{i \to \infty}\ip{\beta \otimes \phi_i}{(\Y\circ \Fsf ^{\Sy})(X)\beta \otimes \phi_i} \\
 & \nonumber = \lim_{i \to \infty}\tr{\Fsf^T(X)\tau_*(P[\beta \otimes \phi_i])}
\end{align*}
for each $X \in \mathcal{B}(S^1)$ and where the limit 
is taken across a set of high amplitude coherent states of the reference. 

The absolute phase $\Fsf^{\Sy}$ can be reconstructed in homodyne detection experiments ({\it e.g.} \cite{psp})
in which the reference system is provided by a local oscillator in a high-amplitude coherent state.
Since $\Fsf^{\Sy}$ is sensitive to the difference between a coherent state $\ket{\beta}$ and ${\tau_{\Sy}}_*(P[\ket{\beta}])$, we may conclude that the pairs $(\ket{\beta}, \phi_i)$ are mutually coherent.
This mutual coherence takes on the appearance of ``absolute" coherence of a laser in the state
$\ket{\beta}$ in the large amplitude limit of the $(\phi_i)$. We stress that the mutually coherent pair
$(P[\ket{\beta}],P[ \phi_i])$ and the mutually incoherent pair $(P[\ket{\beta}],{\tau_{\R}}_* P[\phi_i])$
represent two different physical situations, resulting in different observed statistics in homodyne
experiments. Thus it can be empirically decided that laser light is coherent, though in a 
different sense than is usually discussed, and that a coherent state $P[\ket{\beta}]$ of a laser along with an ``absolute" phase {\sc pom} is an accurate reduced description of the state of affairs, physically different from the description afforded by the state 
${\tau_{\Sy}}_*(P[\ket{\beta}])$.

\section{Concluding Discussion}

We have seen that symmetry necessitates a relational view of states and observables,
the usual textbook ``absolute" description featuring as a convenient shorthand, applicable
where a reference system is suitably localised and may be treated as external. 
``Absolute" quantities and coherence of states
are to be understood as approximate descriptions of relational attributes of system and reference together.

Through experimentation it has been established that in many cases appropriate reference systems
exist and that therefore the ``absolute" description
does prove empirically adequate. ``Absolute" position (as ``phase" conjugate to momentum), angle and optical phase appear to be such instances, and the latter
bears upon the debate on the reality of optical coherence. It is legitimate in computations 
to use ``absolute" phases and coherent states, and these theoretical
notions refer not to the system alone, but only to the combination of one system with another. 

There is an essential physical difference between situations in which ``absolute" quantities
and (``absolutely") coherent superpositions do provide an accurate account of observed statistics
and in which they do not. There is no basis
to expect that localised reference states exist for all phase-like quantities, and therefore the agreement between ``absolute" and relative may not always be exact \cite{mlbshort}. At the extreme
end (complete delocalisation for the reference), only invariant quantities of $\Sy$ (equivalently, states with no coherence) capture the observed statistics. The scenario that ``absolute" quantities not invariant under symmetry do not yield what is observed may well arise for a symmetry
related to particle indistinguishability \cite{frede}; it may be impossible in this case to create appropriate reference states or mutually coherent pairs.

\noindent{\bf Acknowledgements}
Thanks are due to Stephen Bartlett, Dennis Dieks, Chris Fewster, Terry Rudolph and Rob Spekkens  for helpful conversations, and to Rebecca Ronke for valuable feedback on earlier drafts of this manuscript.
TM acknowledges JSPS KAKENHI (grant no. 15K04998). LL acknowledges support under the grant \emph{Quantum Mathematics and Computation} (no.  EP/K015478/1).


\begin{thebibliography}{0}

\bibitem{bor1} N.~Bohr, 
The quantum postulate and the recent development of  atomic theory,
Nature {\bf 121}, 580 (1928).

\bibitem{ed1} A.S.~Eddington, {\it The Nature of the Physical World}, Macmillan (1928).

\bibitem{as} Y.~Aharonov, L.~Susskind, Charge superselection rule, 
Phys.~Rev.~{\bf 155}, 1428 (1967).



\bibitem{ak1} Y.~Aharonov, T.~Kaufherr, Quantum frames of reference, 
in \emph{Proceedings of the International Symposium on Foundations of Quantum Mechanics, Tokyo} (Physical Society of Japan) 1984, pp. 190-194.


\bibitem{ak2} Y.~Aharonov, T.~Kaufherr, Quantum frames of reference,
Phys.~Rev.~D.~{\bf 30}, 368  (1984).


\bibitem{sp1} R.M.~Angelo, N.~Brunner, S.~Popescu, A.J.~Short, P.~Skrzypczyk, Physics within a quantum reference frame,
J.~Phys.~A: Math.~Theor.~{\bf 44}, 145304 (2011).

\bibitem{brs} S.D.~Bartlett, T.~Rudolph, R.~W.~Spekkens, 
Reference frames, superselection rules, and quantum information, 
 Rev.~Mod.~Phys.~{\bf 79}, 555 (2007). 
 

\bibitem{pb1} M.~C.~Palmer, F.~Girelli, S.~D.~Bartlett, Changing quantum reference frames Phys. Rev. A {\bf 89}, 052121 (2014)

 
 \bibitem{www} G.C.~Wick, A.S.~Wightman, E.P.~Wigner, Intrinsic parity of elementary particles,
Phys.~Rev.~{\bf 88}, 101 (1952).

\bibitem {strowi}  F.~Strocchi, A.S.~Wightman, Proof of the charge superselection rule in local relativistic quantum field theory,
J.~Math.~Phys. {\bf 15}, 2198 (1974).



\bibitem{www70} G.C.~Wick, A.S.~Wightman, E.P.~Wigner, 
Superselection rule for charge, 
Phys.~Rev.~D {\bf 1}, 3267 (1970).
 


\bibitem{dia} S.D.~Bartlett, T.~Rudolph, R.W.~Spekkens, Dialogue concerning two views on quantum coherence: factist and fictionist,
Int.~J.~Quantum Inform.~{\bf 4}, 17 (2006).


\bibitem{molmer} K.~M\o lmer,  Optical coherence: a convenient fiction, 
Phys.~Rev.~A {\bf 55}, 3195 (1997).


 
\bibitem{ms1} I.~Marvian, R. W.~Spekkens, 
An information-theoretic account of the Wigner-Araki-Yanase theorem,
arXiv:1212.3378 (2012).

\bibitem{ajr1} M.~Ahmadi, D.~Jennings, T.~Rudolph, The WAY theorem and the quantum resource theory of asymmetry, New~J.~Phys.~{\bf 15} 013057 (2013).

\bibitem{pian1} M.~Piani, M.~Cianciaruso, T.~R.~Bromley,
C.~Napoli, N.~Johnston, G.~Adesso, Robustness of asymmetry and coherence of quantum states,
Phys. Rev. A {\bf 93} (2016) 042107.

\bibitem{pian2} C.~Napoli, T.~R.~Bromley, M.~Cianciaruso, M.~Piani, N.~Johnston, and G.~Adesso,
Robustness of Coherence: An Operational and Observable Measure of Quantum Coherence, Phys. Rev. Lett. {\bf 116}, 150502, (2016).


\bibitem{mlbshort} T.~Miyadera, L.~Loveridge, P.~Busch, 
Approximating relational observables by absolute quantities: a quantum accuracy-size trade-off,
J.~Phys.~A {\bf 49} 185301 (2016).


\bibitem{lbm1} L.~Loveridge, P.~Busch, T.~Miyadera, Symmetry, Reference Frames, and Relational Quantities in Quantum Mechanics,
arXiv:1704.10434 (2017).

\bibitem{hol1} A.S.~Holevo, Generalized imprimitivity systems for Abelian groups,
Sov.~Math.~Iz.~VUZ {\bf 27}, 53 (1983).

\bibitem{ljp1} P.~Lahti, J-P.~Pellonp\"{a}\"{a}, Covariant phase observables in quantum mechanics, J. Math. Phys. {\bf 40}, 10 (1999).

\bibitem{mlak} W.~Mlak, Hilbert Spaces and Operator Theory, Springer (1991).

\bibitem{tnorm1} T.~Heinonen, P.~Lahti, J.-P.~Pellonp\"a\"a, S.~Pulmannova, K.~Ylinen, The norm-1-property of a quantum observable,
J.~Math.~Phys. {\bf 44}, 1998  (2003).


\bibitem{jpthesis} J.-P.~Pellonp{\"a}{\"a},  
{\em Covariant phase observables in quantum mechanics},
PhD Thesis, Annales Universitatis Turkuensis AI 288 (2002), 
\href{http://urn.fi/URN:ISBN:951-29-2123-5}{ http://urn.fi/URN:ISBN:951-29-2123-5}.



\bibitem{pekju}P.~Lahti, J.-P.~Pellonp\"{a}\"{a}, 
Characterizations of the canonical phase observable,
 J.~Math.~Phys. {\bf 41}, 7352 (2000).



\bibitem{dbrs} M.R.~Dowling, S.D.~Bartlett, T.~Rudolph, R.W.~Spekkens, Observing a coherent superposition of an atom and a molecule,
Phys.~Rev.~A {\bf 74}, 052113 (2006).



\bibitem{psp} J.-P.~Pellonp\"{a}\"{a}, J.~Schultz, M.G.A.~Paris,
Balancing efficiencies by squeezing in realistic eight-port homodyne detection,
Phys.~Rev.~A {\bf 83}, 043818 (2011).

\bibitem{frede} K.~Fredenhagen, Quantum Field Theory Lecture notes, available at 
\href{http://unith.desy.de/research}{http://unith.desy.de/research} (2009).


\end{thebibliography}
\end{document}